\documentclass[conference]{IEEEtran}
\IEEEoverridecommandlockouts
\usepackage{cite}
\usepackage{amsmath,amssymb}
\usepackage{graphicx}
\usepackage{booktabs}
\usepackage{url}
\usepackage{tikz}
\usetikzlibrary{arrows.meta,positioning}
\usepackage[T1]{fontenc}

\usepackage{microtype}
\usepackage{enumitem}
\def\BibTeX{{\rm B\kern-.05em{\sc i\kern-.025em b}\kern-.08em T\kern-.1667em\lower.7ex\hbox{E}\kern-.125emX}}

\begin{document}

\title{Trace2ST:\\ Recovering PLC Structured Text from I/O Traces\\ with LLM Generation and Verification}

\author{\IEEEauthorblockN{Hsi Chang}
\IEEEauthorblockA{School of Electrical, Computer and Energy Engineering\\ Arizona State University\\ Phoenix, Arizona, USA\\ xhan570@asu.edu}
\and
\IEEEauthorblockN{Irfan Khan}
\IEEEauthorblockA{Clean and Resilient Energy Systems (CARES) Lab\\ Marine Engineering Technology, in a joint affiliate appointment\\ with Electrical and Computer Engineering\\ Texas A\&M University, College Station, USA\\ irfankhan@tamu.edu}}

\maketitle

\begin{abstract}
Programmable logic controllers (PLCs), the industrial computers that operate power and process plants, often run programs whose source code is no longer available; however, every input and output of the controller is logged continuously. This paper therefore asks whether a large language model (LLM) can recover an executable program in Structured Text (ST), the standard PLC language of IEC 61131-3, from such input/output records (traces) alone and, equally important, how the recovered program can be verified. To answer this question, we build a generate--replay--feedback pipeline. First, the LLM writes a candidate program from a compressed summary of the trace, without ever seeing the original program. The candidate is then executed one control cycle (scan) at a time and compared with the recording, and whenever a mismatch occurs the corresponding part of the trace is returned to the LLM, for at most three revisions. Because every result depends on the software that executes the candidates, we first validate it against two independent open-source PLC runtimes and find no unexplained difference. On this basis, we evaluate 47 third-party programs from two public benchmarks. The recovered programs reproduce the recorded trace exactly in 43 cases and agree with the original on 94\,\% of scans for unseen inputs. The remaining errors, however, lie mostly in program branches that the trace never exercised; once a trace that targets these branches is added, agreement on unseen adversarial inputs rises from 0.80 to 0.98. By contrast, a classical automata-learning baseline without an LLM fits the recording but falls to 0.40 on unseen inputs, which indicates that recovering program structure is the hard part. Finally, on real data from a public water-treatment testbed, the recovered control laws match unseen measurements closely ($R^2 > 0.99$) and expose the only actuator attack in the dataset.
\end{abstract}


\section{Introduction}
Substations, water-treatment plants and generation auxiliaries are operated by PLCs whose control logic is often decades old. Over such a lifetime the source code frequently becomes unavailable, because it is lost, survives only as a compiled image, or is held in a proprietary engineering tool. The plant historian, that is, the data archive of the plant, in contrast remains intact: it stores every input and output of the controller as a time series, typically at one sample per second. If the logic could be recovered from these records, three practical tasks would benefit directly, namely migrating the controller to a new platform, building a digital twin (a software copy of the plant used for simulation), and detecting intrusions. The last task depends on such a model in particular, since knowing what the controller \emph{should} do is the most direct way to recognise a manipulated actuator~\cite{shin2020hai}.

LLMs have recently been applied to PLC programming, but only in the \emph{forward} direction, in which a natural-language requirement is turned into code with compilers and model checkers in the loop~\cite{koziolek2023chatgpt,fakih2024llm4plc,liu2024agents4plc}. This paper studies the \emph{inverse} problem instead. Here neither a requirement nor the source is available, and the trace is the only evidence. The question is therefore whether an LLM can write an ST program that behaves like the original controller.

The main difficulty of this inverse problem, however, is not generation but \emph{verification}, because a recovered program may look plausible and still match the recording only by coincidence. For this reason we build the study around a strict protocol. First, the LLM is physically separated from the reference programs. Second, a candidate is judged only by replaying it against recorded data. Third, every claim is checked on inputs the LLM never saw. Finally, two baselines without an LLM are run on the same data, so that the contribution of the LLM can be isolated.

Following this protocol, the paper makes four contributions.
\begin{itemize}[leftmargin=*,itemsep=1pt,topsep=2pt]
\item \textbf{A trace-to-ST pipeline with a validated judge.} We implement an interpreter, i.e., software that executes ST programs scan by scan on an ordinary computer, and cross-check it against two independent runtimes; as a result, none of the 144 comparisons shows an unexplained difference.
\item \textbf{A blind benchmark.} It contains 47 third-party programs, each with a recorded trace, a held-out trace, adversarial probe traces and specification invariants, and it is complemented by repeated runs.
\item \textbf{A comparison with non-LLM baselines.} Rule mining and passive automata learning fit the recording but do not generalise (0.40 against 0.94 on held-out traces); hence the value of the LLM lies in recovering program structure.
\item \textbf{A case study on real plant data.} On the HAI 23.05 testbed the recovered control laws reach $R^2$ of 0.995 and 0.997 and, in addition, expose an actuator-manipulation attack.
\end{itemize}

\section{Problem Statement and Terminology}
Table~\ref{tab:terms} collects the terms used in this paper. A PLC executes its program cyclically. In each \emph{scan} it reads all inputs, runs the program once, and writes all outputs, while internal state such as timers and counters is carried from one scan to the next. A \emph{trace} is accordingly a table with one row per scan, and to \emph{replay} a program on a trace means to feed it the recorded inputs row by row and to compare its outputs with the recorded ones.

\begin{table}[t]
\centering\caption{Terms used in this paper.}
\label{tab:terms}
\footnotesize
\begin{tabular}{@{}p{2.05cm}p{6.25cm}@{}}
\toprule
PLC & Industrial computer that reads sensors and drives actuators. \\
Scan (cycle) & One pass of the PLC: read inputs, run the program once, write outputs (100\,ms in our experiments). \\
Structured Text & Textual PLC programming language of IEC 61131-3 (ST). \\
I/O trace & Table with one row per scan holding every input and output value. \\
Replay & Feeding the recorded inputs to a program and comparing its outputs with the record. \\
Scan agreement & Fraction of scans on which all outputs equal the record. \\
Interpreter & Our software that executes ST on a PC in place of a PLC. \\
RuSTy & Open-source ST compiler; we use its parser to check that a candidate is valid ST. \\
truST & Open-source ST compiler and runtime that ships a conformance test suite. \\
MATIEC & Open-source compiler that translates ST into C code. \\
OpenPLC & Open-source PLC platform; it uses MATIEC as its compiler. \\
HAI & Public dataset recorded on an industrial testbed with real controllers (boiler, turbine, water treatment). \\
Held-out trace & Trace from randomly generated inputs of the same kind but a different random sequence; never shown to the agent. \\
Probe trace & Trace designed to trigger rarely reached behaviour. \\
Spec.\ invariant & Rule stated by a program's authors: ``whenever $a$ holds, $c$ must hold''. \\
Automata learning & Classical method that builds a state machine from traces. \\
Baseline & A method without an LLM, used for comparison. \\
\bottomrule
\end{tabular}
\end{table}

With these terms, the problem can be stated as follows. Given one or more traces of an unknown program $P$ and its interface (variable names and types), produce an ST program $\hat{P}$ whose outputs equal those of $P$, not only on the given traces but also on inputs that were not recorded. We measure this by \emph{scan agreement}. Because $P$ has internal state, however, the task cannot be solved by learning a function from the current inputs to the current outputs; instead, the program structure, including its timers and state variables, must be inferred.

\section{Related Work}
Two lines of work are closest to this problem. \emph{Forward generation} is the closest in technique: Koziolek et~al.\ evaluated ChatGPT on control-logic prompts~\cite{koziolek2023chatgpt}, and LLM4PLC~\cite{fakih2024llm4plc} and Agents4PLC~\cite{liu2024agents4plc} subsequently added compilers and formal verification to the generation loop. These methods start from a specification, whereas ours starts from observed behaviour; the two are therefore complementary. \emph{Model learning}, in turn, is the closest in goal. Inferring a controller from its behaviour is classically treated as automata learning~\cite{vaandrager2017}: active algorithms (for example the one known as $L^*$~\cite{angluin1987}) query the system, while passive algorithms (for example those known as RPNI~\cite{oncina1992rpni} and EDSM~\cite{lang1998edsm}) build a state machine from recordings by repeatedly merging states that the data cannot tell apart. Since only recordings are available in our setting, we implement the passive family as a baseline.

For execution semantics we follow the IEC standard~\cite{iec61131} and rely on three open-source tools, the parser of RuSTy~\cite{rusty}, which checks that a program is syntactically valid ST, the runtime truST~\cite{trust}, and MATIEC~\cite{tisserant2007matiec}, the compiler behind OpenPLC~\cite{alves2014openplc}; specification invariants in the style of the verification tool PLCverif~\cite{darvas2015plcverif} provide a check that is independent of the traces.

\section{Method}

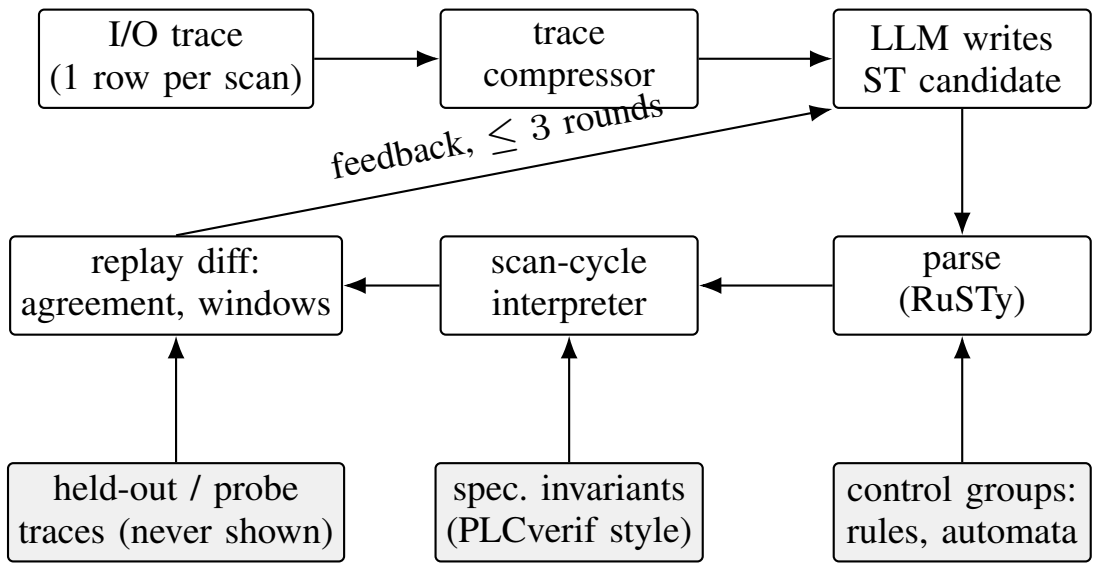
\begin{figure}[t]
\centering
\begin{tikzpicture}[every node/.style={font=\scriptsize},
  box/.style={draw, rounded corners=1pt, align=center, minimum height=6.5mm, minimum width=17mm, inner sep=2pt},
  arr/.style={-{Latex[length=1.5mm]}}]
\node[box] (tr) at (0,0) {I/O trace\\(1 row per scan)};
\node[box] (tc) at (2.6,0) {trace\\compressor};
\node[box] (llm) at (5.2,0) {LLM writes\\ST candidate};
\node[box] (cmp) at (5.2,-1.5) {parse\\(RuSTy)};
\node[box] (eng) at (2.6,-1.5) {scan-cycle\\interpreter};
\node[box] (diff) at (0,-1.5) {replay diff:\\agreement, windows};
\node[box, fill=gray!12] (pr) at (0,-3.0) {held-out / probe\\traces (never shown)};
\node[box, fill=gray!12] (doc) at (2.6,-3.0) {spec. invariants\\(PLCverif style)};
\node[box, fill=gray!12] (ctl) at (5.2,-3.0) {control groups:\\rules, automata};
\draw[arr] (tr) -- (tc); \draw[arr] (tc) -- (llm); \draw[arr] (llm) -- (cmp);
\draw[arr] (cmp) -- (eng); \draw[arr] (eng) -- (diff);
\draw[arr] (diff.north) -- node[above,sloped,pos=0.5]{feedback, $\le 3$ rounds} (llm.south west);
\draw[arr] (pr) -- (diff); \draw[arr] (doc) -- (eng); \draw[arr] (ctl) -- (cmp);
\end{tikzpicture}
\caption{Pipeline. The LLM sees only the compressed trace and the mismatch feedback, never the reference program. Grey boxes are used for evaluation only: unseen traces, specification invariants, and the two non-LLM baselines, which are scored by the same interpreter.}
\label{fig:pipe}
\end{figure}

Fig.~\ref{fig:pipe} shows the pipeline that addresses the problem defined above. A trace is first compressed into a summary; the LLM then writes a candidate program from this summary; next, the candidate is compiled and replayed; and finally the mismatches are returned to the LLM.

\subsection{Trace compression}
The pipeline starts with compression, because a trace of 4\,000 scans is too long to place in a prompt and, moreover, most of its rows repeat the previous one. The compressor therefore reduces the trace to a summary of 7--20\,kB with four parts: (i) a variable table with declared types, value ranges and transition counts; (ii) a sample of output transitions, each with the input values at that scan and the input changes just before it; (iii) the raw rows around rare input and output events; and (iv) automatically mined hints. A hint is a candidate regularity, for example a threshold (``\texttt{K01} rises when \texttt{PIT01}$-$\texttt{SP} $< -0.5$'') or a constant delay between an input change and an output edge, which in turn suggests a timer. So that the LLM can weigh them, each hint is scored by the number of output edges it explains.

\subsection{Candidate generation and feedback}
Given this summary, the LLM (Claude) writes a function block, which is ST's unit of program with its own internal state, with the recorded interface. A new agent with an empty context is started for each program, so that nothing carries over between programs. A driver then compiles the candidate, replays it on every training trace, and writes a feedback file that contains the agreement of each output and up to four mismatch windows of 16 scans. On the basis of this feedback, the agent revises the candidate. To prevent open-ended trial and error, however, each agent may request at most three checks, and a compile error counts as one. In the end, the candidate with the highest training agreement is kept and is scored on traces the agent could not access.

\subsection{Execution engine and its validation}
The replay step above requires software that executes ST programs. Since no physical PLC is involved, candidates are executed by an \emph{interpreter} that we wrote: a program that runs on an ordinary computer and performs, scan by scan, what a PLC would do, with time supplied by a virtual clock. It covers the ST subset used in control code, namely the elementary types with their declared widths, arrays, structures, control statements, the standard timers and counters, and about 60 standard functions.

Since every number in this paper is produced by this interpreter, its correctness had to be established first. We therefore compared it with two independent open-source tools that also execute ST, on the reasoning that independently written programs are unlikely to share the same error. First, truST~\cite{trust} provides a conformance suite, that is, a set of test programs with known correct outputs; the interpreter passes all 11 applicable cases, and five semantic defects were found and corrected in the process. Second, MATIEC~\cite{tisserant2007matiec} translates ST into C code, which we drove scan by scan with the same inputs. As a result, the two agree on every scan in 135 of 144 program--trace pairs. The remaining nine pairs, which come from three programs, differ for a single reason: MATIEC converts between \texttt{TIME} and numbers in seconds whereas our interpreter uses milliseconds, a choice the standard leaves to the vendor. Once the unit is switched, these nine pairs also agree exactly, and consequently no difference remains unexplained. The comparison also exposed an error in how the interpreter handled number widths, which was corrected before the results below were produced.

\subsection{Evaluation metrics}
With a trusted interpreter in place, candidates can be scored. The main metric is scan agreement, where Booleans and integers must match exactly and reals within a relative tolerance of 0.1\,\%. It is reported on three kinds of trace, which test generalisation with increasing strictness. The \emph{recorded} trace is the one the agent works from. The \emph{held-out} trace, by contrast, comes from randomly generated inputs of the same kind but a different random sequence, and is never shown to the agent. \emph{Probe} traces go one step further: they are designed from the interface alone to reach behaviour that random inputs rarely trigger, such as a reset after a fault, a set-point crossed in both directions, or a counter driven to its maximum.

Scan agreement alone can nevertheless be misleading, so two further measures complement it. \emph{Output coverage} is the fraction of outputs that change at least once in a trace; it is needed because an output that never changes is matched trivially. \emph{Specification consistency}, in addition, is the number of invariants written by the benchmark authors that the replayed candidate satisfies on all traces.

\section{Experimental Setup}
\begin{figure}[t]
\centering
\begin{tikzpicture}[x=0.97cm,y=1cm,every node/.style={font=\scriptsize},
  box/.style={draw, rounded corners=1.5pt, align=center, minimum height=7.5mm, inner sep=2pt},
  arr/.style={-{Latex[length=1.5mm]}}]
\node[box, minimum width=50mm] (ref) at (3.5,0) {Reference program (hidden from the agent)\\ 3 synthetic + 47 third-party};
\node[box, minimum width=40mm, fill=gray!12] (rec) at (1.2,-1.45) {Recorded trace\\ random inputs, sequence A};
\node[box, minimum width=19mm] (held) at (4.95,-1.45) {Held-out trace\\ sequence B};
\node[box, minimum width=19mm] (probe) at (7.05,-1.45) {Probe trace\\ designed};
\node[box, minimum width=19mm, fill=gray!12] (agent) at (0.1,-2.85) {LLM agent\\ $\le 3$ checks};
\node[box, minimum width=19mm, fill=gray!12] (base) at (2.3,-2.85) {Baselines: rules,\\ automata};
\node[box, minimum width=40mm, fill=gray!12] (cand) at (1.2,-4.15) {Candidate ST programs};
\node[box, minimum width=84mm] (score) at (3.55,-5.35) {Scoring: replay in the validated interpreter on all three traces; specification invariants};
\draw[arr] (ref.south) -- ++(0,-0.25) -| (rec.north);
\draw[arr] (ref.south) -- ++(0,-0.25) -| (held.north);
\draw[arr] (ref.south) -- ++(0,-0.25) -| (probe.north);
\draw[arr] (rec.south -| agent) -- (agent.north);
\draw[arr] (rec.south -| base) -- (base.north);
\draw[arr] (agent.south) -- (agent.south |- cand.north);
\draw[arr] (base.south) -- (base.south |- cand.north);
\draw[arr] (cand.south) -- (cand.south |- score.north);
\draw[arr, dashed] (held.south) -- (held.south |- score.north);
\draw[arr, dashed] (probe.south) -- (probe.south |- score.north);
\node[align=center] at (6.0,-3.3) {scoring only,\\ never shown};
\end{tikzpicture}
\caption{Experimental setup. Each reference program is executed with three sets of inputs to produce three traces. The agent and the baselines see only the recorded trace (grey); the held-out and probe traces are used for scoring only. In round~2, one probe trace is also given to the agent.}
\label{fig:setup}
\end{figure}
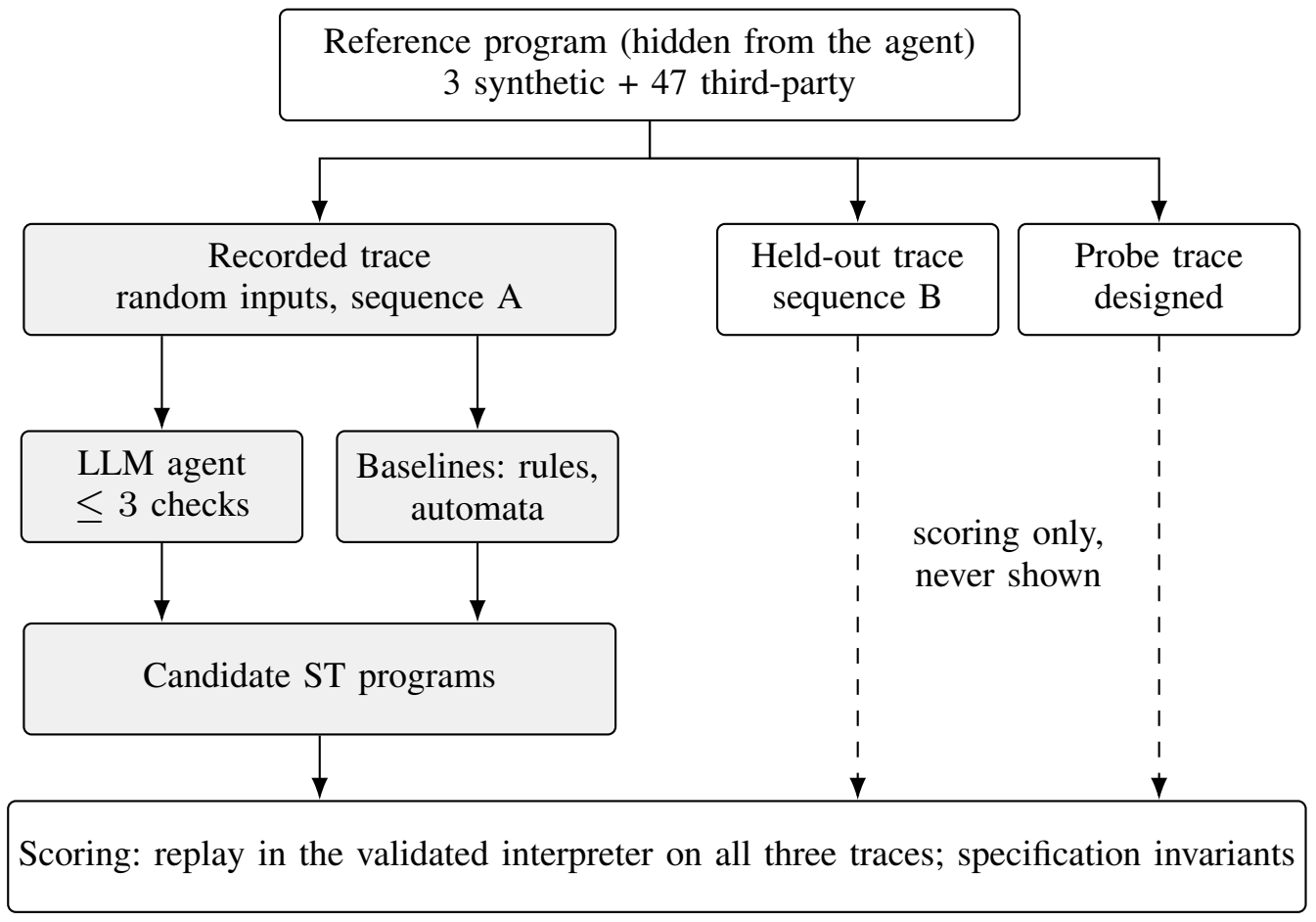

Fig.~\ref{fig:setup} summarises the setup in which the method is evaluated.

\textbf{Programs.} The experiments use two groups of programs. The first group serves to calibrate the tool chain and consists of three synthetic controllers (a tank-level controller, a conveyor and filler sequence, and a pressure controller), each simulated together with the process it controls. The second group is the main benchmark. It comprises 47 programs with simple (non-array) inputs and outputs from two public sources that are independent of the authors: PLCHaT~\cite{plchat} contributes hand-written utility blocks such as an alarm with hysteresis, a pulse generator and a traffic light, whereas the second source~\cite{plcgenstudy} contributes composite interlock programs, which block an action until its safety conditions are met, and which come with requirements and properties. Each trace has 4\,000 scans of 100\,ms. Two programs inspected during development were excluded.

\textbf{Blind protocol.} To keep the reconstruction blind, the reference sources were removed from the machine before every LLM run. Consequently, an agent could read only the prompt, the feedback and the recorded trace. Two early agents nevertheless opened a held-out trace before this rule was enforced, and they are excluded from the held-out results.

\textbf{Baselines.} To isolate what the LLM contributes, two methods without an LLM are applied to the same traces. \emph{Rule mining} selects, for each output, the best rise rule and the best fall rule among the compressor's hints, adds an on-delay timer where a constant delay was found, and emits ST. \emph{Passive automata learning}, in contrast, treats Boolean inputs as symbols and discretises numeric inputs at the thresholds observed at output edges. It then builds a state machine from all training scans with EDSM~\cite{lang1998edsm}; for a state--input pair that was never observed, it returns the most frequent output for that input. Its hyper-parameters are selected on training agreement only. Finally, the learned machine is exported as ST and scored by the same chain as the LLM candidates, so that all methods are compared on equal terms.

\section{Results}
The results are organised as five research questions (RQs), each of which builds on the previous one.

\subsection{RQ1: Can an LLM recover programs from traces?}
\textbf{Finding.} \emph{Yes for most programs: 43 of 47 candidates reproduce the recorded trace exactly, and mean agreement on held-out traces is 0.94.}

As a first step, we ran the pipeline on the three synthetic systems, with the LLM session that developed the pipeline writing the candidates. All three controllers were matched exactly on the recorded traces, and held-out agreement was 1.00, 0.70 and 1.00. The value of 0.70 has an instructive cause: the program contains a timer whose duration the trace bounds only from below, and so the candidate assumed 1\,s where the true value is 1.5\,s. The error thus stems from missing evidence rather than from faulty reasoning, a pattern that recurs below.

\begin{table}[t]
\centering\caption{Blind reconstruction of 47 third-party programs (round 1: recorded trace only, at most three checks, one fresh agent per program).}
\label{tab:r1}
\begin{tabular}{lc}
\toprule
recorded-trace agreement, mean (exact) & 0.965 (43/47) \\
held-out agreement, mean (exact), $n{=}45$ & 0.941 (25) \\
probe agreement, mean (exact) & 0.832 (12) \\
programs with 100\,\% output coverage & 26 (held-out 0.93) \\
rule-mining baseline, recorded & 0.232 \\
\bottomrule
\end{tabular}
\end{table}

Having confirmed that the pipeline works end to end, we ran the blind protocol on the 47 third-party programs; Table~\ref{tab:r1} summarises this first round. Forty candidates matched the recorded trace at the first check and 43 after at most three. On held-out traces, moreover, the mean agreement is 0.941, and it remains 0.93 for the 26 programs in which every output changes. The result is therefore not an artefact of inactive outputs.

\subsection{RQ2: Where does recovery fail, and can it be repaired?}
\textbf{Finding.} \emph{Failures concentrate in branches that the trace never exercised. One additional probe trace raises agreement on unseen probe inputs from 0.80 to 0.98.}

Although most programs are recovered, a residual error of 1--10\,\% remains in some. To locate it, we examined the agents' own notes. They consistently attribute the unexplained scans to behaviour that the trace does not show: a reset with no observable effect, a counter that never reaches its limit, a starter that never starts. The probe traces confirm this explanation. Under the probe inputs, output coverage rises from 0.82 to 0.91, whereas agreement falls to 0.832 (Table~\ref{tab:r1}). As a consequence, some programs that had appeared solved are exposed; a pulse generator, for example, falls from 1.00 to 0.13.

If the cause is indeed missing evidence, then supplying the evidence should repair the candidates. We tested this hypothesis on the 31 programs with imperfect probe agreement. Fresh agents received the recorded trace together with one probe trace, and their candidates were then evaluated on a second probe trace generated with a different random sequence.

\begin{table}[t]
\centering\caption{Effect of one probing round on 31 programs. Both columns are evaluated on a probe trace unseen by either candidate.}
\label{tab:r2}
\begin{tabular}{lcc}
\toprule
 & round 1 & round 2 \\
\midrule
mean agreement on unseen probe & 0.803 & \textbf{0.984} \\
exactly 1.0 & 1 & \textbf{20} \\
$\ge 0.95$ & 14 & \textbf{29} \\
improved / same / worse & \multicolumn{2}{c}{29 / 1 / 1} \\
\bottomrule
\end{tabular}
\end{table}

Table~\ref{tab:r2} shows that the hypothesis holds. Mean agreement rises from 0.803 to 0.984, and 30 of the 31 agents matched both training traces after a single check. The limiting factor in round~1 was therefore the evidence in the trace, not the capability of the LLM.

The same conclusion follows from a check that does not rely on traces at all. The 30 composite programs come with 275 specification invariants, 201 of them related to safety. While the reference programs satisfy 273 of them, round-1 candidates satisfy 250 and round-2 candidates 271; furthermore, 26 of the 30 round-2 candidates satisfy every invariant. The two invariants violated by the reference programs, in turn, are genuine inconsistencies between specification and implementation that the probes exposed: a capacity can be lowered below the current count without clamping, and a high-high limit can be configured below the high limit.

\subsection{RQ3: Is the LLM necessary?}
\textbf{Finding.} \emph{The baselines fit the recording but do not generalise: automata learning reaches 0.97 on the training trace and 0.40 on held-out traces, against 0.94 for the LLM.}

\begin{table}[t]
\centering\caption{Baselines and LLM on the same traces (mean scan agreement).}
\label{tab:ctrl}
\begin{tabular}{lcccc}
\toprule
trace & automata & table & rules & LLM \\
\midrule
recorded (training) & 0.970 & 0.838 & 0.232 & 0.965 \\
held-out ($n{=}45$) & 0.397 & 0.379 & -- & \textbf{0.941} \\
probe ($n{=}47$) & 0.336 & 0.310 & -- & \textbf{0.832} \\
rec.+probe $\to$ new probe ($n{=}31$) & 0.439 & -- & -- & \textbf{0.984} \\
\bottomrule
\end{tabular}
\end{table}

The results so far show that the LLM can recover programs; they do not yet show that an LLM is required. Table~\ref{tab:ctrl} therefore compares it with the two baselines. On the training trace the automata learner is as accurate as the LLM, which is expected because it is constructed from that trace. On unseen traces, however, its agreement falls to 0.397 and 0.336. The last row is the most informative: a second training trace improves the automata learner only from 0.34 to 0.44, whereas it brings the LLM to 0.98. Additional evidence is thus useful only to a learner that can turn it into structure.

The reason lies in how timers and counters are represented. An automaton has no notion of elapsed time; consequently, a 3\,s timer becomes 30 states at a 100\,ms scan, one program produced 5\,512 states, and a delay not present in the training data cannot be represented at all. The LLM, by contrast, expresses the same behaviour as a single timer instance. Accordingly, the automata learner is better only on three programs that the LLM also failed in round~1, where memorised transitions happen to be closer than an incorrect structural hypothesis.

\subsection{RQ4: Are the results robust?}
\textbf{Finding.} \emph{Repeated runs agree, the candidates are accepted by an independent compiler, and two experiments rule out recall of the reference programs.}

The findings above rest largely on single runs of one model, so three threats were examined in turn. The first is \emph{repeatability}. Ten programs (four solved, three intermediate and three failed in round~1) were re-run with three fresh agents each. Of the 30 runs, 29 matched the recorded trace at the first check and the last at the second. Probe agreement, moreover, differed between agents by at most 0.09, and only where the trace leaves a branch undetermined. The three failed programs had been reconstructed in round~1 by the development session rather than by a fresh agent; all nine fresh agents, in contrast, solved them exactly. Although substituting these scores would raise the round-1 held-out mean to 0.968, Table~\ref{tab:r1} is reported unchanged.

The second threat is that the candidates might run only in our interpreter. However, 74 of the 86 best candidates also compile under MATIEC, and all 12 failures are syntax strictness issues such as a missing semicolon, not logic errors.

The third threat is \emph{training-data contamination}: because the benchmarks are public, the model might be recalling the programs rather than inferring them. Two experiments address this concern. In the first, ten programs were replaced by published mutants, that is, copies with one deliberate change in behaviour, while the original names were kept; the mutants differ from the originals on 23--100\,\% of held-out scans. If the model were recalling, it would reproduce the originals. Instead, every candidate follows the mutant: mean held-out agreement is 0.981 against the mutant's trace but only 0.409 against the original's. In the second experiment, all names were replaced by neutral identifiers. Held-out agreement was then 0.966 against 1.000 with names, and probe agreement 0.825 against 0.831. Names therefore help marginally, yet the method does not depend on them.

\subsection{RQ5: Does the method transfer to real plant data?}
\textbf{Finding.} \emph{A blind run on the HAI testbed recovers the pump and valve laws with $R^2$ of 0.995 and 0.997 on unseen data, and the recovered program exposes the dataset's actuator attack.}

All previous experiments use traces generated by executing programs, which are clean and complete. To test the method on real measurements, we therefore turn to HAI~\cite{shin2020hai,hai2305}, which records a hardware-in-the-loop testbed, that is, real controllers connected to a simulated plant, at 1\,Hz. Its water-treatment process P3 is controlled by a Siemens S7-300, a widely used industrial PLC, with two analog actuators, a level-control valve and a pump, both driven by the load demand of a power-plant simulation. The exported load variables and the tank level serve as inputs. Since the outputs are continuous, we report the coefficient of determination $R^2$ between replayed and recorded commands instead of scan agreement; a value of 1 means that the two curves coincide.

As a reference, we first analysed the data by hand. The pump law consists of two hysteresis gates, each of which switches off above one threshold and on again only below a lower one (total load above 352\,MW, or level above 69.6\,\%), followed by a first-order lag, so that the command approaches its target gradually with a time constant of 20\,s (roughly the time needed to cover two thirds of the way), while the valve is a linear function of the load whose speed of change is capped. This pump law reaches $R^2 = 0.992$ on the 78-hour training set and, without refitting, 0.991 on the 64-hour test set.

We then repeated the task blind. To this end, all analysis notes were removed, and a fresh agent received a 24-hour slice under the same protocol. It nevertheless arrived independently at the same pump structure, with a lag coefficient of 0.050 and gates at 346 and 352\,MW, and in addition at a more accurate valve law. On the unseen test set, its program consequently reaches $R^2 = 0.997$ for the valve and 0.995 for the pump, compared with 0.984 and 0.992 for the hand-derived laws.

\begin{figure}[t]
\centering\includegraphics[width=\columnwidth]{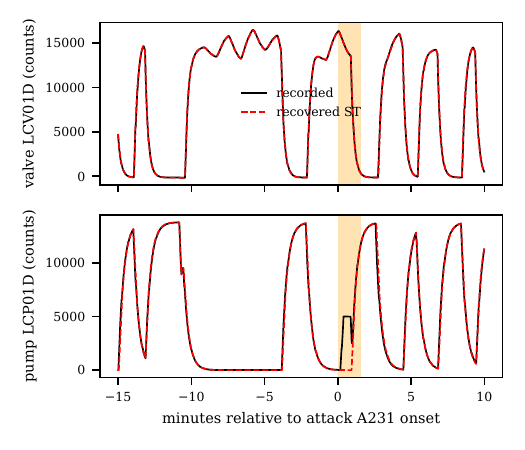}
\caption{HAI test set around attack A231 (shaded): recorded actuator commands and the commands computed by the blind-recovered program from the same inputs. The two coincide in normal operation; the largest deviation is the attack itself, during which the pump is driven while the recovered law keeps it off.}
\label{fig:hai}
\end{figure}

Beyond reproducing normal operation, the recovered program can be used directly for intrusion detection (Fig.~\ref{fig:hai}). The underlying idea is that the residual, i.e., the difference between the recorded command and the command computed by the program, is small in normal operation but grows when an actuator is manipulated. The test set contains one such attack on P3, in which the pump command is manipulated for 96\,s. Its residual is the eighth largest among 3\,839 one-minute windows (fourteenth with the hand-derived law). Accordingly, a simple rule, a residual above 3\,000 counts for 10\,s, detects the attack with 12 false alarms in 64 hours, whereas a fixed normal range on the raw signal produces 489. All 12 false alarms, moreover, occur when the load is within $\pm 1$\,MW of the recovered gate, which suggests that the controller uses a filtered or unexported signal at that point.

\section{Threats to Validity}
\emph{Single model family.} All agents, and the session that developed the pipeline, use the same LLM family, so the findings have not been confirmed with a second vendor. \emph{Benchmark.} The 30 composite programs share an interlock skeleton and thus represent roughly 5--8 distinct families; moreover, random inputs leave at least one output inactive in 21 of 47 programs, which is why we report coverage and add the probe round. \emph{Single trials.} Most cells are single runs, although the repeat study bounds the spread between agents at 0.09. \emph{Baseline strength.} The automata learner is textbook EDSM, not a learner for automata with timers; the comparison therefore supports the claim that structure recovery is hard, not that automata learning is ineffective. \emph{Data.} The true inputs of the HAI controller are not exported and hence were reconstructed, and vendor-defined semantics such as time units must be known for the target runtime.

\section{Conclusion}
This paper asked whether an LLM can recover a PLC program from input/output traces alone. The results show that it can, provided that two conditions are met. The first is a trustworthy judge of correctness: candidates must be judged by replay, and the replay engine itself must be validated independently, which we regard as a prerequisite for any claim of this kind. The second is sufficient evidence: on 47 third-party programs nearly all errors lie in branches the trace never exercised, and accordingly one round of targeted probing raises agreement on unseen probe inputs from 0.80 to 0.98. Under these conditions, the recovered programs are executable and readable, generalise to unseen inputs, and transfer to real plant data, whereas non-LLM learners reproduce the recording without capturing the structure behind it. Building on these findings, future work will generate probes automatically from the agents' notes on unexplained behaviour, extend trace compression to continuous processes, and repeat the evaluation with a second LLM vendor.

\section*{Acknowledgment}
The software, traces and candidates are released with this paper. Generative AI (Claude, Anthropic) was used in this work at three levels. (i) It is the subject of the study: all candidate programs were written by Claude agents, as described in Sections IV--VI. (ii) The software used in the experiments, including the interpreter, the trace compressor, the baselines and the evaluation scripts, was developed with Claude under the authors' direction, and the experiments were executed with its assistance. (iii) The text, tables and figures were prepared with Claude under the authors' dictation and outline, and were then reviewed and revised by the authors. The authors verified the results and take full responsibility for the content of this paper.

\bibliographystyle{IEEEtran}
\bibliography{refs}

\end{document}